# Unifying temperature definition in atomistic and field representations of conservation laws

Youping Chen
Department of Mechanical and Aerospace Engineering, University of Florida, Gainesville, Florida 32611

## ABSTRACT

This work presents a formalism to derive field quantities and conservation laws from the atomistic using the theory of distributions as the mathematical tool. By defining temperature as a derived quantity as that in molecular kinetic theory and atomistic simulations, a field representation of the conservation law of linear momentum is derived and expressed in terms of temperature field, leading to a unified atomistic and continuum description of temperature and a new conservation equation of linear momentum that, supplemented by an interatomic potential, completely governs thermal and mechanical processes across scales from the atomic to the continuum. The conservation equation can be used to solve atomistic trajectories for systems at finite temperatures, as well as the evolution of field quantities in space and time, with atomic or multiscale resolution. Four sets of numerical examples are presented to demonstrate the efficacy of the formulation in capturing the effect of temperature or thermal fluctuations, including phonon density of states, thermally activated dislocation motion, dislocation formation during epitaxial processes, and attenuation of longitudinal acoustic waves as a result of their interaction with thermal phonons.

## I.   INTRODUCTION

Temperature is a basic concept in the broad field of physical science. It is a fundamental physical quantity in thermodynamics, statistical mechanics, and continuum mechanics. Specifically, the principal variable in thermodynamics is the absolute temperature, which is introduced as a basic quantity in the zeroth law. In statistical mechanics of canonical ensemble, temperature serves as a basic parameter characterizing the heat reservoir with which the system contacts. In classical continuum mechanics, the axiom of causality states that "the motion, temperature, and charges of the material points of a body are the cause of all physical phenomena"[1,2], thereby prescribing temperature as one of the basic quantities upon which all other quantities are derived.

Different from the macroscopic or continuum field theories that treat temperature as a basic quantity, kinetic molecular theory expresses temperature as a derived quantity based on classical mechanics of particles. Built on the work of Bernoulli[3], Clausius derived a relationship between temperature and the average kinetic energy of gaseous systems[4]. This kinetic energy-based temperature, generally referred to as kinetic temperature, has been the abiding temperature definition in all classical atomistic simulations.

While temperature is a ubiquitous quantity in theoretical formulations of physical theories[5,6], it is also a quantity that is frequently measured in experiments. The engineering concept of temperature can be traced back to 1593 when Galileo Galilei invented a thermoscope to measure hotness or coldness based on the thermal expansion of air[7]. One hundred years later, Medici invented a thermometer, which was later improved by Amontons, by defining temperature in terms of the pressure of a gas. Building on the works on the thermal expansion of liquids by Romer in 1708 and Fahrenheit in 1724[8,9], Celsius introduced a scale of 100 degrees for the temperature interval between the ice point and the boiling points of water[10]. There are other thermometers, most of which were developed based on a material property that depends on temperature. Thus, temperature in most experimental measurements is a derived quantity.

Despite its ubiquity and fundamental importance, there is no unified definition of temperature among well-established theories, making it difficult for multiscale simulations that involve different theoretical descriptions. This is especially true for concurrent atomistic-continuum simulations, as temperature is a derived quantity in mechanics of particles but a basic quantity in mechanics of continua; the former is completely governed by Newton's second law, whereas for the latter, one must solve both the conservation equation of linear momentum and the conservation equation of energy assuming a heat flux-temperature relation. This fundamental inconsistency limits the applicability of concurrent multiscale methods for simulation of nonequilibrium materials processes, constituting one most formidable challenge for the development of multiscale methods.

This work aims to address the challenge by deriving a unified description of temperature in atomistic and continuum representation of conservation laws. Following the introduction, microscopic descriptions of temperature are reviewed in Sec. II, including the kinetic molecular theory description of temperature and pressure, and the phonon description of temperature. The concurrent atomistic-continuum (CAC) formulation of the field representation of conservation laws is then reviewed in Sec. III. In Sec. IV we derive both averaged and instantaneous conservation equations of linear momentum in terms of temperature field. In Sec. V we present a finite element implementation of the formulation. Sec. VI presents four numerical examples to demonstrate the efficacy of the formulation. This paper concludes with a summary and discussions in Sec.VI.

## II. A REVIEW OF MICROSCOPIC DESCRIPTIONS OF TEMERATURE

### 2.1 Temperature in molecular kinetic theory

Attempts to derive a mathematical description for pressure and temperature from microscopically moving particles can be traced back to 1729 when Euler formulated a rudimentary kinetic theory[11] based on Boyle's gas data published in 1662[12]. Using the assumption that all the gas particles move at the same speed, Euler obtained an equation of state connecting $P$ (pressure), $\rho$ (density), and $v$ (molecular speed) for dry air, i.e., $P \propto \rho v^2 /3$. In 1733, Bernoulli published his book on hydrodynamics[13], in which he derived an equation for the pressure of a gas considering point-like particles of mass colliding with the walls of the container and showed that the pressure is two thirds of the average kinetic energy of the gas in a unit volume. Built on the atomic theory that a gas consists of moving particles, Clausius derived the relationship between temperature and averaged kinetic energy of gas particles in thermal equilibrium in his paper "On the Nature of the Motion which We Call Heat" in 1857[4] as

$$\left\langle \frac{1}{2}m\boldsymbol{v}_k^2 \right\rangle = \frac{3}{2}k_B T \;, \tag{1}$$

where $\boldsymbol{v}_k$ is the velocity vector of a particle $k$, $<>$ denotes time average, and $k_B$ was later called Boltzmann constant[14]. Eq.(1) has been widely used as a measure of temperature for equilibrium systems. In the molecular kinetic theory and molecular dynamics (MD) simulations, it is assumed that Eq.(1) remains valid in the absence of thermal equilibrium and the definition then coincides with that indicated by a sufficiently small and fast thermometer.

Clausius's kinetic theory was further developed by Maxwell based on a statistical viewpoint. Considering that the collisions among molecules would produce a statistical distribution of molecular velocities, Maxwell developed a distribution law for the velocities, known as Maxwell's Distribution, providing the first mathematical tool to find the statistical velocity distribution of molecules of an ideal gas in a box at a definite temperature $T$, thereby giving temperature a statistical meaning. Building on the works of Clausius and Maxwell, Boltzmann formulated a statistical theory that is not restricted to ideal gases and showed that the distribution function in any system at thermal equilibrium at temperature $T$ has the canonical form:

$$f(E) \propto \exp\left(-\frac{E}{k_B T}\right) \;, \tag{2}$$

where $E$ is the energy of the system. Applying this distribution to the velocity of a molecule in an ideal gas, one obtains the Maxwell's velocity distribution, as well as the temperature in Eq.(1) but as an ensemble average.

### 2.2 Pressure in the molecular kinetic theory

The molecular kinetic theory assumes that pressure of a gaseous system in a container is caused by the force associated with individual atoms striking the walls of the container. Consider a particle $k\alpha$ ($\alpha$-th particle in the $k$-th unit cell) traveling with velocity $\boldsymbol{v}_{k\alpha}$ and striking a wall of the container, i.e., a face of the cube, as illustrated in Fig.1. The force exerted at the wall due to the collision is equal to the change of the linear momentum of the particle during the collision. For example, the z-component of collision force can be expressed based on Newton's second law as

$$F^z = \frac{\Delta p_{k\alpha}^z}{\Delta t} = \frac{mv_{k\alpha}^z - (-mv_{k\alpha}^z)}{\Delta t} = \frac{2mv_{k\alpha}^z}{\Delta t} \;. \tag{3}$$

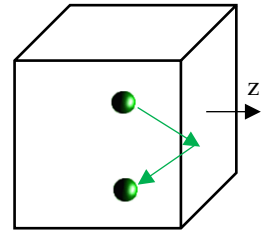

Fig.1 A particle striking a wall and bouncing back.

In thermal equilibrium, time- averaged pressure of the system is uniform, and the formula can then be expressed in the form of a volumetric density. Denote $V$ to be the volume of a cube with edge length $L$. The average time for a molecule to hit the wall is $2L/v_{k\alpha}^z$. The contribution from particle $k\alpha$ to the pressure can then be expressed as

$$P_{k\alpha}^z = \frac{F^z}{L^2} = \frac{2mv_{k\alpha}^z}{\Delta t L^2} = \frac{m(v_{k\alpha}^z)^2}{L^3} = \frac{\frac{1}{3}m[(v_{k\alpha}^x)^2 + (v_{k\alpha}^y)^2 + (v_{k\alpha}^z)^2]}{V}, \quad (4)$$

which, according to Eq. (1) is $k_B T/V$. For a system with N particles, this then yields the ideal gas law,

$$PV = Nk_B T. \quad (5)$$

Thus, pressure is closely related to temperature, and in Sec. IV we will show that the local pressure in Eq. (4) is identical to the kinetic or thermal stress.

## 2.3 Phonon description of temperature

Different from the classical concept of temperature, the quantum description of temperature is in terms of phonons. A phonon is a quantum mechanical description of atomic vibrations, in which a lattice of atoms collectively oscillates at a single frequency. There are more phonons at higher temperature and fewer phonons at lower temperature. The average number of phonons with wavevector $\kappa$ and polarization $v$ at temperature $T$ is given by

$$n(\mathbf{k},v) = n(\omega, T) = 1 / \left[ \exp\left(\frac{\hbar\omega(\mathbf{k},v)}{k_B T}\right) - 1 \right], \quad (6)$$

where $\hbar$ is the Planck constant and $k_B$ the Boltzmann constant. This relationship is called the Planck distribution or the Bose-Einstein distribution. The total kinetic energy of the system, which is equal to the potential energy at thermal equilibrium, can then be expressed in terms of all the available phonons modes in the systems as

$$K = \frac{1}{2} \sum_{k,v} \hbar\omega(\mathbf{k},v) \left[\frac{1}{2} + n(\mathbf{k},v)\right]. \quad (7)$$

At high temperature all phonon modes have the same energy; half of this energy is kinetic which is equal to $k_B T/2$. The phonon description of temperature then becomes analogous to the classical particle description of temperature where each degree of freedom has an average kinetic energy of $k_B T/2$. Since the number of the phonon modes is equal to the number of degrees of freedom, for systems at temperature higher than the Debye temperature, the total kinetic energy of the system in Eq. (7) then becomes

$$K = \frac{1}{2} \sum_{k=1}^{N_l} \sum_{v=1}^{3N_a} \hbar\omega(\mathbf{k},v) \left[\frac{1}{2} + n(\mathbf{k},v)\right] = \frac{1}{2} \sum_{k=1}^{N_l} \sum_{\xi=1}^{N_a} m_{k\xi}(v_{k\xi})^2 = \frac{3}{2} N_l N_a k_B T. \quad (8)$$

The difference between quantum and classical temperatures can be quantified through comparing the total kinetic energy of phonons to that of atoms. Thus, for a given quantum temperature, the corresponding classical temperature can be calculated[15]. Fig. 2 presents such calculations for single crystal Fe, Si, and SrTiO3, which show that the two temperatures are different at low temperature, but the difference becomes negligible as temperature increases.

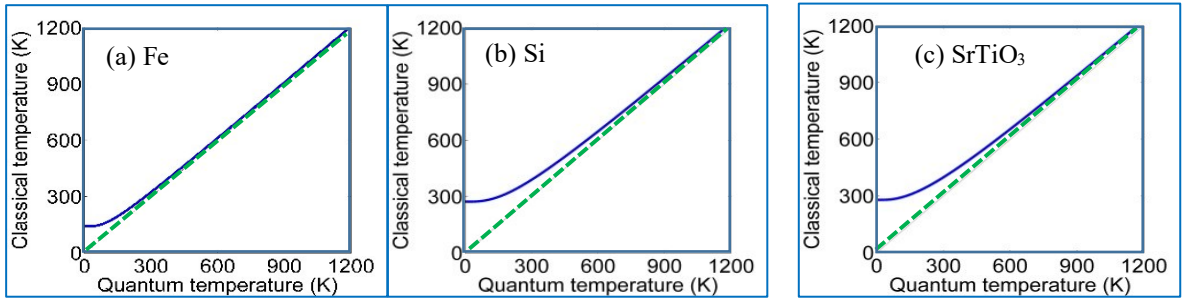

Fig. 2 Quantum (green dashed line) vs classical temperature (blue solid curves) for (a) Fe, (b) Si, and (c) SrTiO3, showing that the two temperatures converge at high temperature. Note that at 0 K quantum temperature, Fe, Si, and SrTiO3 have a zero-point energy equivalent to 142.7K, 273.7K, 279.7K, respectively.

# III. A REVIEW OF CAC FORMULATION OF CONSERVATION LAWS

The first derivation of the field equations of conservation laws from a molecular model was published in 1950 by Irving and Kirkwood[16], who used the Dirac delta to define the local density of a dynamic function $A(r_k,v_k)$ in phase space as an ensemble-averaged point function $\bar{a}(x,t)$ in the physical space, as illustrated in Fig. 3, as

$$\bar{a}(x,t) \equiv \left\langle \sum_{k=1}^{N} A(r_k,v_k)\delta(x-r_k) \right\rangle , \qquad (9)$$

where the bracket $\langle\ \rangle$ denotes ensemble average (meaning repeating an experiment many times), $r_k$ and $v_k$ are the position and velocity of $k^{th}$ molecule, and the $\delta$-function is defined by its sifting property[17], i.e.,

$$\iiint_\Omega f(x)\delta(a-x)d^3x = f(a) . \qquad (10)$$

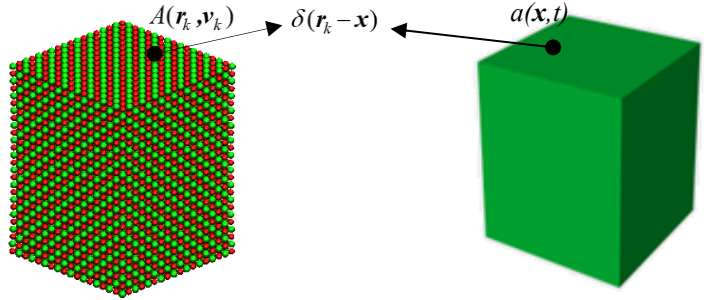

Fig. 3 Mapping a phase function $A(r_k,v_k)$ of a particle $k$ to the local density of $A$ at point $x$ in the physical space.

Irving and Kirkwood then derived the time evolution equations for the densities of mass, momentum, and energy, and related the rate of change of the ensembled averaged local density of a conserved quantity, $\bar{a}(x,t)$, to its ensemble-averaged flux, $\bar{J}(x,t)$, in the following form

$$\frac{\partial}{\partial t}\bar{a}(x,t) = \nabla_x \cdot \bar{J}(x,t) . \qquad (11)$$

Equation (11) is identical in form to the equations of hydrodynamics for homogenized continua. It was noted by Irving and Kirkwood (IK) that all the quantities in their formulation were defined as point functions and "to obtain the hydrodynamical equations themselves it is necessary to perform appropriate space and time averages"[16], and that their formulation "was only valid for single-phase single-component systems". In the first one of a series of 14 papers on statistical mechanical theory of transport processes, Kirkwood envisioned the extension of the formulation to "molecular systems with internal degrees of freedom"[18].

The Concurrent Atomistic-Continuum (CAC) formulation[19-21] is such an extension. CAC views a crystal as a continuous collection of lattice cells, while embedded within each lattice cell there is a group of discrete atoms. A similar two-level structural description was developed in the Micromorphic theory[22-25]. By including the internal atomic degrees of freedom (DOFs) within each lattice cell, CAC extends the IK formulation to a concurrent two-level structural description of general crystals. Recently, the CAC conservation equations have been reformulated[19] using the Theory of Distributions[26], which enables CAC equations and flux formulas to be valid instantaneously. In distinction from the IK formulation that define all field quantities as ensemble-averaged point-functions, CAC defines the local densities of mass, linear momentum, and energy as time-interval-averaged volume densities, while fluxes, such as stress and heat flux, are obtained as a time-interval-averaged surface density[19,27-30]. As a result of continuous distribution of unit cells in crystals, the volume densities of mass, linear momentum, and energy are continuous in space in equilibrium and steady-state systems, while the fluxes are valid at the atomic or larger scales without the need of being continuous.

With the two-level structural description of a crystal, CAC expresses the position of an atom as a sum of the position of the lattice cell, $x$, and the internal position of the atom relative to the lattice, $y$, cf. Fig. 4. Correspondingly, the local density of a phase function $A$ per unit-cell volume can be defined as

$$a(x,t) \triangleq \sum_{k=1}^{N_l} \sum_{\alpha=1}^{N_a} A(r_{k\alpha}, v_{k\alpha}) \bar{\delta}_V(x-r_k) \triangleq \sum_{\alpha=1}^{N_a} a_\alpha(x,t) , \qquad (12)$$

where $N_l$ is the number of lattice cells in the system; $N_a$ is the number of atoms within a lattice cell; and $a_\alpha(x,t)$ is the contribution of the $\alpha^{th}$ atom to $a(x, t)$, and can be expressed as

$$a_\alpha(x,t) \triangleq \sum_{k=1}^{N_l} A(r_{k\alpha}, v_{k\alpha}) \bar{\delta}_V(x-r_k) = \sum_{k=1}^{N_l} A(r_{k\alpha}, v_{k\alpha}) \bar{\delta}_V(x+y-r_{k\alpha}) \triangleq a'(x,y,t) , \qquad (13)$$

or as

$$a_\alpha(x,t) = \sum_{\xi=1}^{N_a} \sum_{k=1}^{N_l} A(r_{k\xi}, v_{k\xi}) \bar{\delta}_V(x-r_k) \bar{\delta}_{V_\alpha}(y - \Delta r_{k\xi}) , \qquad (14)$$

where the box functions

$$\bar{\delta}_V(x-r_k) = \frac{1}{\Delta t} \int_t^{t+\Delta t} \delta_V(x-r_k(\tau))d\tau , \text{ where } \delta_V(x-r_k) = \frac{1}{V}\begin{cases} 1 & \text{if } x-r_k \in V \\ 0, & \text{if } x-r_k \notin V \end{cases}, \qquad (15)$$

$$\bar{\delta}_{V_\alpha}(y-\Delta r_{k\xi}) = \frac{1}{\Delta t} \int_t^{t+\Delta t} \delta_{V_\alpha}(y-\Delta r_{k\xi}(\tau))d\tau , \text{ where } \delta_{V_\alpha}(y-\Delta r_{k\xi}) = \begin{cases} 1 & \text{if } y-\Delta r_{k\xi} \in V_\alpha \text{ or } \xi=\alpha \\ 0 & \text{if } y-\Delta r_{k\xi} \notin V_\alpha \text{ or } \xi \neq \alpha \end{cases}, \qquad (16)$$

and $V$ and $V_\alpha$ are the volume of the unit cell and of atom $\alpha$ within the unit cell, respectively, as shown in Fig. 5.

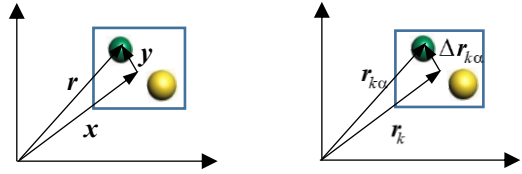

Fig. 4 Left: Position vector, $x$, of a unit cell, and the internal position of atom $\alpha$ relative to the unit cell, $y$, in physical space. Right: the corresponding positions in the phase space.

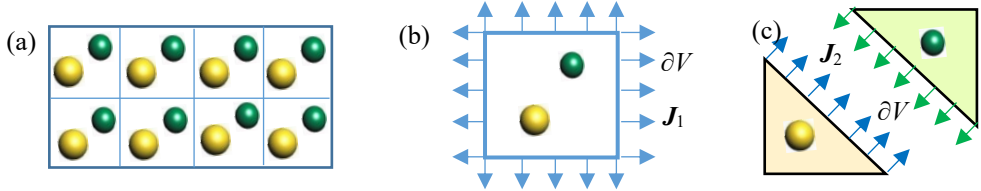

Fig. 5 2D Schematic of (a) a crystal lattice; (b) the flux across the bounding surface of a unit cell $\partial V$ (the blue rectangle), $J_1$; and (c) the flux that does not cross $\partial V$ but $\partial V_\alpha$ (the black triangles) within the unit cell, $J_2$.

Following from Eq.(14), the distributional derivatives of the density of a conserved quantity, i.e., the conservation equation, can be derived and expressed in terms of two fluxes $J_1$ and $J_2$ as[19]

$$\frac{\partial}{\partial t} a_\alpha(x, t) = \frac{\partial}{\partial t} a'(x,y, t) = -\frac{1}{V} \oiint_{\partial V} J_1(x+x', t) \cdot n \, d^2 x' - \frac{1}{V} \oiint_{\partial V_\alpha} J_2(x, y+y', t) \cdot n \, d^2 y' , \qquad (17)$$

where $n$ is the outward unit normal to the differential surface element; $J_1$ is the homogeneous flux that flows across the enclosing surface of a unit cell, $\partial V$; $J_2$ is the inhomogeneous flux that flows back and forth within $V$ and hence only cross the bounding surface of the atom $\alpha$, $\partial V_\alpha$, as illustrated in Fig. 5.

It is noted that a time-interval average is used in the definition of field quantities, by which the fluxes are obtained as a local property averaged over a time interval and a surface element, i.e., as an triple integral of the Dirac $\delta$ [28]. This is consistent with the definition of the $\delta$ by Paul Dirac through integration[17], as the Dirac $\delta$ does not itself have a definite value and only triple integrals of a 3D Dirac $\delta$ are well-defined through the sifting property expressed in Eq.(10). Using such a definition of local densities, fluxes across a surface element are represented as a line-plane intersection process, with the potential part of a flux involving a line of force intersecting a surface element, while for the kinetic part it is the path of a moving particle intersecting a surface element during a time interval [28]. In the theory of distributions, the restriction is lifted. However, to describe the process of the flow of momentum or energy across a surface element, a time-interval average of the local property remains necessary.

It follows from Eq.(14) that the linear momentum density due to the contribution from atom α can be expressed as

$$\rho_\alpha v_\alpha \equiv \sum_{k=1}^{N_l} \sum_{\xi=1}^{N_a} m_\xi v_{k\xi} \bar{\delta}_V(x-r_k) \bar{\delta}_{V_\alpha}(y-\Delta r_{k\xi}). \tag{18}$$

The distributional derivative of the linear momentum density can then be obtained in the form of Eq.(17) as[19]

$$\frac{\partial(\rho_\alpha v_\alpha)}{\partial t} = \frac{1}{V}\oiint_{\partial V}(t_\alpha - \rho_\alpha v_\alpha v \cdot n)dS + \frac{1}{V}\oiint_{\partial V_\alpha}(\tau_\alpha - \rho_\alpha v_\alpha \Delta v_\alpha \cdot n)dS_\alpha, \tag{19}$$

where $t_\alpha$ and $\tau_\alpha$ are the homogeneous and inhomogeneous parts of stress, respectively, with each further containing a kinetic and a potential part. The kinetic part of stress was found to be the flow of momentum per unit area per unit time across a surface element; while the potential stress is the force transmitted per unit area across the surface element and is identical to the Cauchy stress vector. Please see [19] for detailed derivations of the conservation laws, including the energy equation, as well as discussions of above equations and quantities.

## IV. TEMPERATURE IN THE INSTANTANEOUS LINEAR MOMENUM EQUATION

### A. Temperature in averaged linear momentum equation

Alternatively, we can use the local density defined in Eq.(13) and derive a different but mathematically equivalent form of the distributional derivative of the linear momentum density, i.e.,

$$\frac{\partial(\rho_\alpha v_\alpha)}{\partial t} = \sum_{k=1}^{N_l} m_\alpha \dot{v}_{k\alpha} \bar{\delta}_V(r-r_{k\alpha}) + \left(\sum_{k=1}^{N_l} m_\alpha v_{k\alpha} \frac{\partial}{\partial t} \bar{\delta}_V(r-r_{k\alpha})\right) = f^{int}(r) - \frac{1}{V}\oiint_{\partial V}\left(\sum_{k=1}^{N_l} m_\alpha \tilde{v}_{k\alpha} \tilde{v}_{k\alpha} \bar{\delta}(r+r'-r_{k\alpha}) + \rho_\alpha v_\alpha v_\alpha\right) \cdot n\, d^2r', \tag{20}$$

or

$$\rho_\alpha \dot{v}_\alpha = f^{int}(r,t) - \frac{1}{V}\oiint_{\partial V}\sum_{k=1}^{N_l} m_\alpha \tilde{v}_{k\alpha} \tilde{v}_{k\alpha} \bar{\delta}(r+r'-r_{k\alpha}) \cdot n\, d^2r' \triangleq f^{int}(r,t) + f^T(r,t). \tag{21}$$

where $\tilde{v}_{k\alpha}$ is the difference between the particle velocity and the velocity field, and since it vanishes upon time or ensemble averaging, the time average of the tensor product $\tilde{v}_{k\alpha}$ and $v_\alpha$ also vanishes. The relation between averaged kinetic energy density and temperature field can then be expressed, according to Eq. (1), as

$$\left\langle \sum_{k=1}^{N_l} \frac{1}{2} m_\alpha (v_{k\alpha})^2 \delta_V(r-r_{k\alpha}) \right\rangle = \frac{1}{2}\rho_\alpha(v_\alpha)^2 + \left\langle \iiint_V \sum_{k=1}^{N_l} \frac{1}{2} m_\alpha (\tilde{v}_{k\alpha})^2 \delta_V(r-r_{k\alpha}) \right\rangle = k_{\alpha 1}(r,t) + k_{\alpha 2}(r,t) = \frac{3}{2}k_B T(r,t). \tag{22}$$

It follows from Eq.(21) and Eq.(22) that

$$\langle f^T(r,t)\rangle = -\frac{1}{V}\left\langle \oiint_{\partial V}\sum_{k=1}^{N_l} m_\alpha \tilde{v}_{k\alpha} \tilde{v}_{k\alpha} \bar{\delta}(r+r'-r_{k\alpha}) \cdot n\, d^2r'\right\rangle = -\oiint_{\partial V}\left\langle \sum_{k=1}^{N_l} \rho_\alpha(\tilde{v}_{k\alpha})^2 \bar{\delta}(r+r'-r_{k\alpha})\right\rangle n\, d^2r' = -2\oiint_{\partial V}\langle k_{\alpha 2}(r+r',t)\rangle n\, d^2r' = -\beta\nabla T(r,t), \tag{23}$$

where $k_{a1}$ and $k_{a2}$ are the kinetic energy density due to the velocity field $v_\alpha$ and that due to the velocity difference $\tilde{v}_{k\alpha}$, respectively; the bracket < > represents a time or ensemble average by which the averaged function becomes a smooth function; $\beta$ is a constant to be determined by the resolution of the measurement, such as the size of a finite element, and will be discussed in detail in Sec V.

### B. Fluctuating force in coarse-grained models that omit molecular degrees of freedom

As indicated by Eq.(23), the instantaneous $f^T(r,t)$ is a fluctuating function with its average being proportional to the temperature gradient. Since the fluctuating force is an irregular function in time, one may attempt to express it as a random function. It turns out that this fluctuating force cannot be simply expressed as a random force, but a combination of a random force and a frictional force. The underlying principle was first demonstrated by Einstein who applied statistical methods to the random motions of Newtonian atoms and formulated a "general molecular theory of heat"[31], and explained the random movement of pollen particles immersed in a liquid observed by botanist Robert Brown in his experiments, leading to the development of the theory of Brownian motion. Over the past decades, the theory has been further developed and extended to systems beyond Brownian particles, such as serving as a coarse-grained model that accounts for omitted molecular degrees of freedom.

The fundamental equation in the theory of Brownian motion is the Langevin equation that contains both a frictional force and a fluctuation force in the equation of motion. To demonstrate the need of a friction force, let us consider a simplified equation of motion with constant mass density $\rho$ under a fluctuating force with density $\boldsymbol{\eta}(t)$, i.e.,

$$\rho \dot{\boldsymbol{v}} = \boldsymbol{\eta}(t) \quad \text{or} \quad \rho \dot{v}_i = \eta_i(t), \tag{24}$$

where $v_i$ and $\eta_i$ are the $i$-th ($i$ = 1, 2, 3) component of the velocity field and the fluctuation force, respectively. The fluctuation force is generally assumed to be not correlated with the velocity $v_i(0)$. If we assume the fluctuation force is delta-function correlated, i.e.,

$$\langle \boldsymbol{\eta}(t) \rangle = 0, \; \langle \eta_i(t) \eta_j(t') \rangle = 2g \delta_{ij}(t-t'), \quad \text{where} \; \delta_{ij}(t-t') = \begin{cases} 1 \; \text{if} \; i=j \; \& \; t=t' \\ 0 \; \text{otherwise} \end{cases}, \tag{25}$$

Equation (24) can be solved to show that the mean velocity vanishes but the kinetic energy increases with time, as

$$\langle v_i \rangle = \int_0^t \left\langle \frac{\eta_i(t)}{\rho} \right\rangle dt = 0 \quad \text{and} \quad \left\langle \frac{1}{2} \rho v_i^2 \right\rangle = \frac{1}{2} \rho \int_0^t dt \int_0^t \left\langle \frac{\eta_i(t)}{\rho} \cdot \frac{\eta_i(t')}{\rho} \right\rangle dt' = \frac{g}{\rho} t . \tag{26}$$

Adding a viscous damping force, $\gamma \boldsymbol{v}$, to Eq.(24), the equation of motion becomes $\rho \dot{v}_i + \gamma v_i = \eta_i(t)$, and the solution for the velocity can then be found as

$$v_i(t) = v_i(0) e^{-\lambda t/\rho} + \int_0^t \frac{\eta(\tau)}{\rho} e^{-\gamma(t-\tau)/\rho} d\tau . \tag{27}$$

As $t$ increases, the mean velocity goes to zero, but the average kinetic energy remains finite, as

$$\left\langle \frac{1}{2} \rho v_i^2 \right\rangle = \frac{1}{2} \rho \int_0^t d\tau \int_0^t \left\langle \frac{\eta_i(\tau) e^{-\gamma(t-\tau)/\rho}}{\rho} \cdot \frac{\eta_i(\tau') e^{-\gamma(t-\tau')/\rho}}{\rho} \right\rangle d\tau' = \frac{g}{2\gamma} . \tag{28}$$

This means the modified equation of motion is able to reproduce the thermal equilibrium state of the systems after a sufficiently long time. Using the temperature definition in Eq.(1) to express the kinetic energy per DOF as $k_B T/2$, or the kinetic energy density as $k_B T/2V$, where $V=m/\rho$ is the volume of the unit cell or the particle at point $\boldsymbol{x}$, we obtain from Eq.(28) the relationship between the strength of the fluctuation, $g$, the friction coefficient, $\gamma$, and temperature at thermal equilibrium, $T$, as

$$g = \gamma k_B T / V \quad \text{or} \quad gV = \gamma k_B T . \tag{29}$$

Equation (29) shows that the frictional force and the random force are related and must be present simultaneously[32], which is an important result of the fluctuating-dissipation theorem for systems. The nature of the random force is generally assumed to be independent of the presence of the force field[32], and the theorem has been shown to remain valid when a potential field is added to the equation of motion[32]. For systems in nonequilibrium steady state, it has been shown that the results obtained for thermal equilibrium systems can still be valid if the velocity is replaced with the relative velocity with respect to the local mean velocity"[33].

The classical Langevin equation has been generalized to non-Markovian processes, for which the fluctuating force does not have to be delta-function correlated but retains long or finite time memory. Consequently, the friction force becomes a memory kernel. Eq. (24) then becomes:

$$\rho \dot{v}_i + \int_0^t \gamma(t-\tau) v_i(\tau) d\tau = f_i^{\text{int}}(\boldsymbol{r},t) + \eta_i(t) , \tag{30}$$

where $\gamma(t-\tau)$ is called dissipative memory kernel, and $\eta_i(t)$ is generally taken to be Gaussian, i.e.,

$$\langle\eta(t)\rangle=0, \quad \langle\eta_i(t)\eta_j(\tau)\rangle=\delta_{ij}C(t-\tau), \text{ and } \quad C(t-\tau)=\frac{2g}{\sigma\sqrt{2\pi}}\exp[-\frac{(t-\tau)^2}{2\sigma^2}], \quad (31)$$

σ is the standard deviation, and 2g is the strength of $\boldsymbol{\eta}(t)$. To ensure that the system achieves an equilibrium or steady state at long times, the dissipative kernel $\gamma(t-\tau)$ and the correlation $C(t-\tau)$ must be related as[34]

$$C(t-\tau)=k_B T\gamma(t-\tau). \quad (32)$$

where $T$ is the temperature of the surrounding heat bath and the relation is assumed to be independent of the internal force field[32,35].

## C. Temperature in the instantaneous linear momentum equation

The instantaneous linear momentum equation, Eq. (21), can thus be rewritten in terms of temperature field as

$$\rho_\alpha \dot{\boldsymbol{v}}_\alpha = \boldsymbol{f}^{\text{int}}(\boldsymbol{r},t)+\boldsymbol{f}^T(\boldsymbol{r},t)=\boldsymbol{f}^{\text{int}}(\boldsymbol{r},t)-\beta\nabla T+\boldsymbol{\eta}(t)-\gamma\boldsymbol{v}_\alpha(\boldsymbol{r},t), \quad (33)$$

if $\boldsymbol{\eta}(t)$ is a delta-correlated white noise shown in Eq. (25), or as

$$\rho_\alpha \dot{\boldsymbol{v}}_\alpha = \boldsymbol{f}^{\text{int}}(\boldsymbol{r},t)+\boldsymbol{f}^T(\boldsymbol{r},t)=\boldsymbol{f}^{\text{int}}(\boldsymbol{r},t)-\beta\nabla T+\boldsymbol{\eta}(t)-\int_0^t\gamma(t-\tau)\boldsymbol{v}_\alpha(\tau)d\tau, \quad (34)$$

if $\boldsymbol{\eta}(t)$ is a time-correlated Gaussian process shown in Eq. (31).

For either case, the random force $\boldsymbol{\eta}(t)$ approximately represents the deviation of instantaneous kinetic energy $k_{\alpha 2}$ from the mean kinetic energy $\langle k_{\alpha 2}\rangle$ due to velocity fluctuations and can be expressed, according to Eq. (23), as

$$\boldsymbol{\eta}(t)=-2\oiint_{\partial V}(k_{a2}-\langle k_{a2}\rangle)\boldsymbol{n}d^2r' \text{ and } \langle\boldsymbol{\eta}(t)\rangle=0. \quad (35)$$

The frictional force then accounts for the difference between $\boldsymbol{f}^T(\boldsymbol{r},t)$ and $-\beta\nabla T+\boldsymbol{\eta}(t)$, which enables the system to reach thermal equilibrium at sufficiently long time and reproduce the kinetic energy-temperature relation.

A few important results and implications of the derivations of this section may be summarized as follows.
(1) Same as that in the IK formulation[16], both the energy and momentum conservation equations in the CAC formulation can be directly derived from Newton's second law[19]. This means that the energy equation is redundant in solving the motion of the systems. Supplemented by a force field $\boldsymbol{f}^{\text{int}}(\boldsymbol{x})$, the linear momentum equations can be used to solve atomic trajectories or to predict the evolution of local densities and fluxes in space and time. The derivation of energy equation is thus not presented in this work. Nevertheless, the energy equation can be used to find the formula of heat flux. Please see references [[19,27-29,36]] for detailed derivations, discussions, and numerical examples.
(2) Equation (19) is an exact representation of conservation law of linear momentum for atomistic systems. It can be used to find formulas to calculate fluxes (stress and heat flux) in atomistic simulations. It also demonstrates that, different from that in monatomic crystals, fluxes in polyatomic systems consist of two parts: a homogenous part that flows across unit cells and an inhomogeneous part that flows back and forth within each unit cell. Each of these fluxes further contains a potential part and a kinetic part. In particular, the kinetic stress, when averaging over certain time-interval, is equal to the negative pressure in molecular kinetic theory presented in Eq. (4).[27]
(3) Equation (21) is also an exact representation of linear momentum conservation law for atomistic systems. It can be solved with atomic-scale accuracy either with the finest finite element mesh or for systems at very low temperature when phonon-phonon scattering can be ignored. Since the equation is expressed in terms of the internal force and the thermal stress, it facilitates finite element implementation, as in nonlinear dynamic finite element methods, the governing equation needs to be expressed in terms of an internal force vector and an external force vector.
(4) Equation (33) or (34) is the linear momentum equation expressed in terms of temperature field. It is an instantaneous equation and can be used to solve systems and processes at finite temperatures with atomic resolution. It is noted that, due to the statistical nature of temperature, the equation represents atomistic systems in the statistical sense.

# V. FINITE ELEMENT IMPLEMENTATION

## 4.1 Kinetic energy and velocity field

While kinetic energy can be uniquely defined, its measurement depends on computational or experimental resolution. For example, in an FE representation, the motion of traditional finite elements only represents a part of the total kinetic energy. To elucidate this point, we use the decomposition of the kinetic energy in Eq. (22) again:

$$\frac{1}{2}\sum_{k=1}^{N_l}\sum_{\xi=1}^{N_a} m_\xi (v_{k\xi})^2 \bar{\delta}_V(x-r_k)\bar{\delta}_{V\alpha}(y-\Delta r_{k\xi}) = \frac{1}{2}\rho_\alpha(v_\alpha)^2 + \frac{1}{2}\sum_{k=1}^{N_l}\sum_{\xi=1}^{N_a} m_\xi (\tilde{v}_{k\xi})^2 \bar{\delta}_V(x-r_k)\bar{\delta}_{V\alpha}(y-\Delta r_{k\xi}) = k_{\alpha 1}(x,t) + k_{\alpha 2}(x,t) . \quad (36)$$

Here, $k_{\alpha 1}$ is the kinetic energy density of α-th atom due to the motion of finite elements, i.e., the FE nodal velocities, $k_{\alpha 2}$ is the kinetic energy due to the velocity difference between particle velocity and the velocity field and is the kinetic energy that would be omitted in a usual FE representation.

From the phonon viewpoint, the motion of finite elements only represents a subset of the phonons of a system[37,38], as $k_{\alpha 1}$ and $k_{\alpha 2}$ are the kinetic energy density from phonons whose wavelengths are longer and shorter than the finite element size, respectively. Nevertheless, the omitted kinetic energy, i.e., $k_{\alpha 2}$, can be calculated based on its relation to $k_{\alpha 1}$. Specifically, the equipartition theorem in kinetic theory states that every DOF of an equilibrium system has the same kinetic energy $k_B T/2$; the ratio of the omitted DOFs to the DOFs of the finite element thus determines the ratio of $k_{\alpha 1}$ to $k_{\alpha 2}$. Alternatively, the Bose-Einstein phonon distribution also provides a relationship between $k_{\alpha 1}$ and $k_{\alpha 2}$.

Away from thermal equilibrium, the particle velocity in Eq. (1) is usually replaced by $\tilde{v}_{k\alpha}$, i.e., the difference between particle velocity and the velocity field. Such a temperature definition is consistent with the physical picture of an ideal-gas thermometer in which the velocity is measured relative to the co-moving frame of the kinetic thermometer[39]. However, this nonequilibrium temperature definition has unambiguous meaning only for steady state systems. This is because, although the definition of total kinetic energy is unambiguous, the velocity field or local mean velocity can only be uniquely defined when it is zero (as in thermal equilibrium) or a constant (as in steady state, as it depends on length and time scales, i.e., the averaging, of the linear momentum. Thus, regardless of whether it is in an atomistic or a multiscale simulation, or whether using the FE method or other methods, temperature in coarse grained models can only be well defined when the velocity field can be ambiguously defined.

## 4.2 Kinetic energy in finite element representation

As shown in Eq.(36), the decomposition of the total kinetic energy into $k_{\alpha 1}$ and $k_{\alpha 2}$ depend on the length scale of the velocity field. In particular, in a FE model, $k_{\alpha 1}$ is represented by the FE nodal velocities, in which the displacement field in an 8-node-element is interpolated with FE shape functions $\Phi_\xi(x)$, i.e.,

$$\hat{u}_\alpha(x,t) = \sum_{\xi=1}^{8} \Phi_\xi(x) U_{\xi\alpha}(t) \triangleq \Phi_\xi(x) U_{\xi\alpha}(t) , \quad (37)$$

where $U_{\xi\alpha}(t)$ is the displacement vector of α[th] atom embedded within the ξ[th] node of the element; the contribution of the α[th] atom to the kinetic energy of an element of volume $V_e$ can thus be expressed as

$$\iiint_{V_e} k_{\alpha 1} dV = \iiint_{V_e} \frac{1}{2}\rho_\alpha (\dot{u}_\alpha)^2 dV = \iiint_{V_e} \frac{1}{2}\rho_\alpha (\sum_{\xi=1}^{8} \Phi_\xi(x) \dot{U}_{\xi\alpha})^2 dV . \quad (38)$$

The kinetic energy of an element can also be expressed in terms of phonons using the phonon dispersion relations of the FE model[37]. If the usual tri-linear shape functions are used, the FE model then cuts off phonons whose wavelengths are smaller than the element size. This means that the FE mesh, i.e., the FE displacement approximation, determines the decomposition of total kinetic energy into $k_{\alpha 1}$ and $k_{\alpha 2}$. There are two methods to determine the decomposition of total kinetic energy into $k_{\alpha 1}$ and $k_{\alpha 2}$.

The first method is based on the classical molecular kinetic theory. For an equilibrium system at temperature higher than the Debye temperature, each phonon or particle DOF has the same kinetic energy $k_B T/2$. The time-interval-averaged kinetic energy of an 8-node hexahedral element containing $n_l$ unit cells is thus related to temperature as

$$\left\langle \iiint_{V_e} k_{\alpha 1} dV + \iiint_{V_e} k_{\alpha 2} dV \right\rangle = n_l \times \frac{3}{2} k_B T \quad . \tag{39}$$

Since the ratio of the DOFs of the finite elements and that of a corresponding atomically resolved model is $8/n_l$, the equipartition theorem gives:

$$\left\langle \iiint_{V_e} k_{\alpha 1}(\mathbf{x},t) dV \right\rangle = \left\langle \iiint_{V_e} \frac{1}{2} \rho_\alpha (\sum_{\xi=1}^{8} (\Phi_\xi(\mathbf{x}) \dot{U}_{\xi\alpha})^2 dV \right\rangle = 8(\frac{3}{2} k_B T) , \tag{40}$$

and

$$\left\langle \iiint_{V_e} k_{\alpha 2}(\mathbf{x},t) dV \right\rangle = \left\langle \iiint_{V_e} \frac{1}{2} \sum_{k=1}^{N_l} m_\alpha (\tilde{v}_{k\alpha})^2 \bar{\delta}_V (\mathbf{x}-\mathbf{r}_k) dV \right\rangle = (n_l-8)(\frac{3}{2} k_B T) . \tag{41}$$

The parameter $\beta$ in Eq. (23) and Eq. (33) is thus equal to $3(n_l-8)k_B/V_e$ for system that is discretized via 8-node hexahedral elements. Eqs. (39)-(41) can be used to prescribe FE nodal velocities for a given initial equilibrium temperature $T$ or to find $k_{\alpha 2}$ from $k_{\alpha 1}$. For elements with the finest mesh, i.e., when there are only 8 unit-cells in each element, the total kinetic energy is identical to the kinetic energy of the corresponding atomically resolved model.

The second method estimates the kinetic energy of the omitted phonons using the phonon description of temperature. The cut-off wavevector $\mathbf{k}_c$ by the finite element shape functions can be identified by comparing the phonon dispersion relations of the finite element model with the corresponding atomically resolved model. The corresponding frequency, $\omega_i$, for each phonon branch, $v$, can be found from the phonon dispersion relations. The phonon number for each omitted phonon mode, $n(k_i, v_i)$ can be calculated using Eq.(6). The kinetic energy of the omitted phonons and that of the element can then be expressed, respectively, as[40,41]

$$\iiint_{V_e} \bar{k}_{\alpha 2}(\mathbf{x},t) dV = \left\langle \iiint_{V_e} \frac{1}{2} \sum_{k=1}^{N_l} m_\alpha (\tilde{v}_{k\alpha})^2 \bar{\delta}_V (\mathbf{x}-\mathbf{r}_k) dV \right\rangle = \frac{1}{2} \sum_{k_i>k_c} \sum_{v_i}^{3N_a} \hbar \omega_i \left[\frac{1}{2} + n(k_i, v_i)\right], \tag{42}$$

and

$$\iiint_{V_e} \bar{k}_{\alpha 1}(\mathbf{x},t) dV = \left\langle \iiint_{V_e} \frac{1}{2} \rho_\alpha (\sum_{\xi=1}^{8} (\Phi_\xi(\mathbf{x}) \dot{U}_{\xi\alpha})^2 dV \right\rangle = \frac{1}{2} \sum_{k_i<k_c} \sum_{v_i}^{3N_a} \hbar \omega_i \left[\frac{1}{2} + n(k_i, v_i)\right]. \tag{43}$$

### 4.3 Balance equation of linear momentum in finite element representation

Equation (19), or Eq. (21), is an exact atomistic field representation of the conservation law of linear momentum that is valid instantaneously at the atomic scale. Eq. (33) is also an atomistic field representation of the conservation law of linear momentum but is expressed in terms of temperature field; it thus only holds statistically. There are two different methods to modify a traditional FE method to solve the linear momentum equation at different temperatures.

### A. Systems at low temperature

At low temperature when phonon-phonon scattering is negligible, Eq.(21) can be solved using a modified finite element method. The velocity field can be modelled with the atomic-scale resolution, and consequently $\mathbf{f}^T(\mathbf{r},t)=0$ at every point in space and time, by replacing the usual FE tri-linear shape functions with those that combine tri-linear shape functions and Bloch wave functions, the latter of which can be constructed in terms of the omitted phonons. The displacement field within each finite element can then be approximated as

$$\hat{\mathbf{u}}_\alpha(\mathbf{x},t) = \sum_{\xi=1}^{8} \Phi_\xi(\mathbf{x})[U_{\xi\alpha}(t) - \mathbf{u}_\alpha^{sh}(\mathbf{x}_\xi, t)] + \mathbf{u}_\alpha^{sh}(\mathbf{x},t) \quad . \tag{44}$$

Here, $\Phi_\xi(\mathbf{x})$ ($\xi$=1, 2, ….8 for 8-node 3D elements) is the conventional FE trilinear shape functions, $U_{\xi\alpha}(t)$ is the displacement vector for $\alpha^{th}$ atom at the $\xi^{th}$ FE node, $\mathbf{x}_\xi$ is the position vector of the $\xi^{th}$ FE node, $\mathbf{u}_\alpha^{sh}(\mathbf{x}_\xi,t)$ is the contributions of all the omitted short wavelength phonons to the displacements of the $\alpha$-th atom at the $\xi$-th node, and

$$\mathbf{u}_\alpha^{sh}(\mathbf{x},t) = \frac{1}{\sqrt{N_l m_\alpha}} \sum_{\mathbf{k},v(\mathbf{k}>\mathbf{k}_c)} Q(\mathbf{k},v) \mathbf{e}_\alpha(\mathbf{k},v) \exp[i(\mathbf{k}\cdot\mathbf{x} - \omega(\mathbf{k},v)t)] , \tag{45}$$

where $e_\alpha(k,v)$ is the eigenvector or the polarization vector that determines the direction in which each atom moves, $\omega(k,v)$ is the frequency, and $Q(k,v)$ is the amplitude of the phonon mode($k$, $v$) and is given by

$$\langle |Q(k,v)|^2 \rangle = \frac{\hbar}{\omega(k,v)}\left[n(\omega,T)+\frac{1}{2}\right] \quad . \tag{46}$$

For concurrent atomistic-continuum (CAC) simulations, the mode-specific amplitude can also be obtained from analysis of the atomic displacements in the atomically resolved region within a CAC model, for which one must take the 3D Fourier Transform of the system one atom in the unit cell at a time (please see [42] for details of the procedure).

The method (a) is applicable for low-temperature simulations of the propagation of ultrashort phonon pulses or scattering on interfaces or dislocations, in which the critically regions can be discretized with atomic resolution, while the perfect single crystalline region can be modelled with finite elements[43]. The transport of short-wavelength phonons within each finite element is thus ballistic. By including short-wavelength phonons in the shape functions, this method has been found to enable waves from an atomic region to pass through a finite element region, with the coherence of the waves being preserved in space and time[42].

## B. Systems at finite temperatures

At higher temperature when short-wavelength phonons are dominated by thermally resistive processes[44], the fluctuations due to the omitted short-wavelength phonons can be modelled as a body force with a constant mean but randomly fluctuating in time with the frequencies of the omitted phonons. The atomic motion is thus random, and the transport of short-wavelength phonons is diffusive within each element. The governing equation for such system is Eq.(33) or Eq. (34), both can be numerically solved using the finite element method, and Eq. (33) may be considered a special case of Eq. (34).

Approximating the displacement field of α-th atom in element $V_e$ as $u_\alpha(x) = \Phi_\eta(x)U_{\eta\alpha}$, the weak form of Eq. (34) can be written as

$$\left(\iiint_{V_e} \rho_\alpha \Phi_\xi \Phi_\eta dV\right)\ddot{U}_{\eta\alpha} + \left(\iiint_{V_e} \rho_\alpha \Phi_\xi \Phi_\eta \int_{t-t_c}^{t} \gamma(t-\tau)\dot{U}_{\eta\alpha}(\tau)d\tau dV\right) = \iiint_{V_e}\Phi_\xi f_\alpha^{int}dV - \iiint_{V_e}\Phi_\xi(\beta\nabla T + \eta(t))dV \quad . \tag{47}$$

The integrals in Eq. (47) can be evaluated using Gaussian quadrature. For elements employing tri-linear shape functions, the temperature gradient in the last term is either zero for thermal equilibrium or a constant for non-equilibrium, implying that the system is modeled as steady state within each finite element. Thus, no numerical integration is needed for the last integral, as $\nabla T + \eta(t)$ does not vary within a finite element. It is noticed that the memory integral in (34) for the friction force is replaced by an integral over a finite time interval. This is because that the friction kernel is an autocorrelation function of $\eta(t)$ and that the correlation time determines the decay time of the friction kernel. The memory integral can thus be replaced by an integral over a finite time interval.[45]

A Gaussian process, as shown in in Eq.(31), can be completed characterized by the width of distribution σ and the strength 2g. For the finite element implementation, σ can be chosen based on the omitted frequencies and the discrete time points[45], and 2g can be determined based on Eq.(35) and the well-established relation of fluctuation of kinetic energy, $\Delta K$, and the temperature in classical systems[45,46], i.e.,

$$\Delta K = \frac{1}{2}\sqrt{k_B T^2 C_v} \quad , \tag{48}$$

The widely studied trigonometric representation for Gaussian processes[45] can be used to generate the random force $\eta(t)$ due to the short-wavelength phonons that are omitted by FE tri-linear shape functions, i.e.,

$$\eta_i(t) = \sum_{n=0}^{N_c-1}[A_n\cos(\omega_n t)+B_n\cos(\omega_n t)] \quad , \tag{49}$$

where $A_n$ and $B_n$ are mutually independent Gaussian random variables having zero mean and variance $C(0)$ such that $<A_n B_m> = 0$ and $<A_n A_m> = <B_n B_m> = C(0)\delta_{nm}$; $N_c$ is the total number of omitted phonons in each element, $\omega_n$ are the frequencies of the omitted phonons, and $A_n$ and $B_n$ are mutually independent Gaussian random variables[35].

## VI. NUMERICAL EXAMPLES

To extend the CAC method for finite temperature processes, Equation (47) has been tested though implementation in the CAC codes using a modified finite element method with tri-linear shape functions[47-49]. To test the formulation and numerical implementations, we build four sets of computer models that are discretized with 8-node rhombohedral elements, with each element containing 8×8×8 primitive unit cells and the surfaces of the elements corresponding to the slip and/or the cleavage planes of the crystals.

### A. Phonon density of state

Phonon density of states (or vibrational density of states) is defined as the number of phonon modes per unit frequency per unit volume. It is a fundamental characteristic of a solid, from which many dynamic materials properties can be determined, and can be computed through calculation of the power spectrum of the mass weighted velocity correlation function[40]. To test the accuracy and efficiency of the formulation and numerical implementation, single crystal models that contain 1,024,000 Si and Fe atoms, respectively, are simulated under constant temperature 300K using both CAC and MD. The CAC models are discretized with uniformly meshed finite elements, each of which contains 512 unit cells. The number of degrees of freedom of the CAC models is thus 1.56% of that of the corresponding MD models. Phonon density of state (PDOS) is then computed from Fourier transform of the velocity autocorrelation functions and averaged over 10 simulations with different initial conditions.

In Fig. 6, we compare CAC and MD results of phonon density of state (PDOS) for Si and Fe single crystals, respectively. As can be seen from Fig. (10), the PDOS of the finite-sized Si and Fe can be reproduced by CAC, in good agreement with MD. This means that all the phonon frequencies that are captured in the MD simulations are present in the CAC simulations. For more verification examples and discussions please see [49].

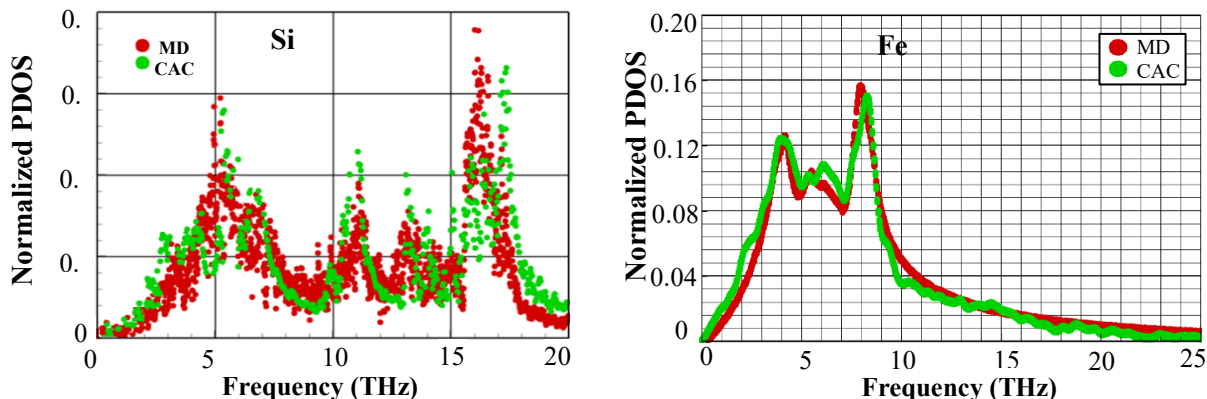

Fig.6 MD and CAC results of normalized PDOS of a Si crystal model (Left) and an Fe crystal (Right) with 512unit cells in each FE. They are computed from the Fourier transform of the velocity autocorrelation functions at 300K.

### B. Thermally activated dislocation motion in bcc iron

Plastic deformation in crystalline materials occurs principally by the motion of dislocations, which is influenced by temperature in two distinct ways: through thermal fluctuations to activate or aide slow dislocation motion, and through phonon scattering to slow down fast dislocation motion. While MD can in principle simulate both processes, temperature-dependent dislocation motion in high-Peierls-stress materials, such as BCC metals, involves much larger length and time scales than that can be reached by a MD simulation[50]. It is well known that the Peierls stress for screw dislocations in BCC iron is high, the motion of dislocations occurs with the aid of the thermally activated kink nucleation and migration to overcome the high-Peierls barriers, and that the mobility of a dislocation depends on the applied stress and temperature, as well as the line length of the dislocations.

To test and demonstrate the formulation on capturing the effect of temperature on the motion of high high-Peierls-stress dislocations, a CAC model of BCC iron that contains a ½<111> screw dislocation is built. The dislocation is introduced into the model through initially displacing the finite element nodes according to the displacement field of a dislocation derived from the theory of elasticity. The computer model has dimensions x=65 nm,

y= 2968 nm and z=16 nm and is discretized into 450,000 finite elements, with the six surfaces of each element being the (110) slip planes of the dislocations. The dislocation has a line length of ~3 μm oriented along y direction and the slip plane is perpendicular to z direction. Periodic boundary condition is applied along y axis. The model is under a constant shear stress of 1.51 GPa and at an equilibrium temperature of 300K. A MD model with the same dimensions would contain ~230 million atoms and hence MD simulation is not conducted, as it is challenging to simulate with our available computational resources. The goal of the simulations is therefore to demonstrate the effect of the thermal fluctuations on dislocation motion.

Figure 7(a) and (b) present snapshots of the dislocation line morphology during the process of dislocation motion and local stress distribution in the vicinity of the dislocation line. The atomic displacements are mapped from the FE nodal displacements. It is seen from Fig.7a that the simulation has captured the two types of dislocation line roughening mechanisms, (1) self-pinning, and (2) debris loop formation. For comparison, we present in Fig.7b the results of a simulation that ignored the short-wavelength phonons. As can be seen from Fig. 7b, neither mechanism has been captured by the simulation without consideration of the short-wavelength phonons.

Both simulations reveal that the dislocation lines propagate through the formation of kink pair. However only the simulation that includes thermal fluctuations due to short-wavelength phonons has captured the cross slip. It is noticed that both self-pinning and the debris loop formation are triggered by cross slips of the dislocation line, as shown in Fig. 8. Note that cross-kinking and debris formation are largely suppressed in (b) as a result of coarse graining that ignores the short-wavelength vibrations. For more simulation results and discussions please see [48,49].

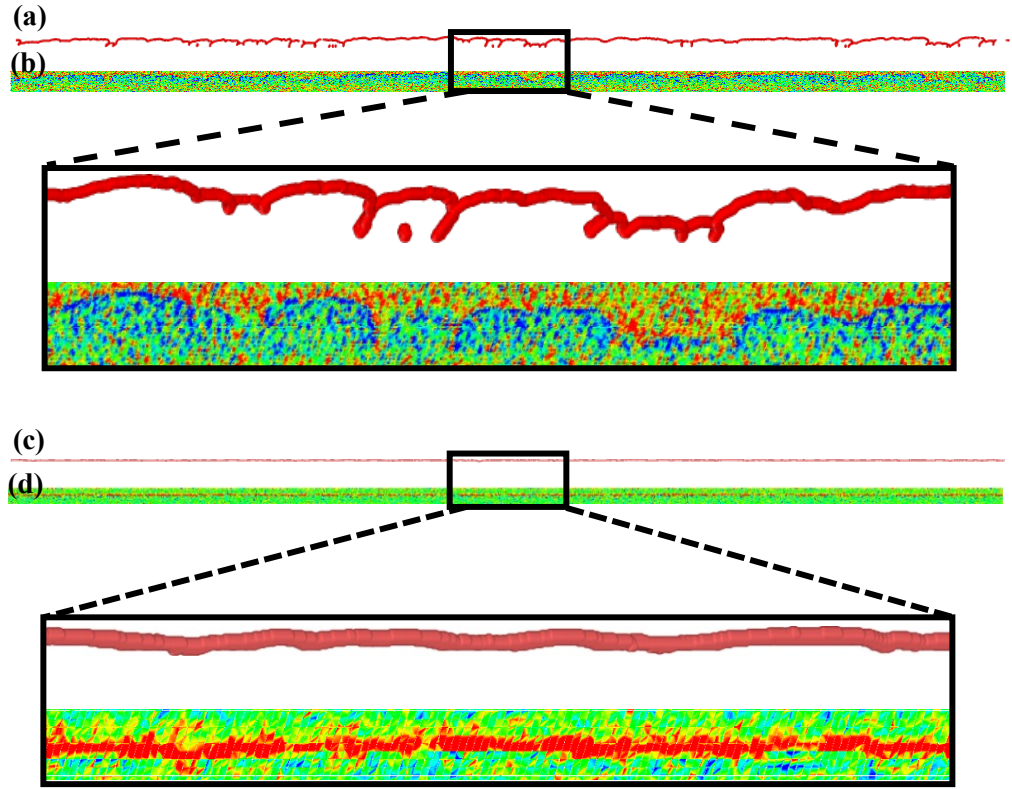

Fig. 7 (a) and (c) snapshots of a 2968-nm-long dislocation line, visualized using common neighbor analysis, showing the morphology of the dislocation line during the motion of the dislocation; (b) and (d) shear stress distribution in the region that contains the dislocation line, where red and blue colors represent positive and negative shear stress, respectively. The atomic positions are mapped from finite element nodal displacements obtained in CAC simulation (a) and (b) using Eq.(47), i.e., with the short wavelength phonons being including in the simulation; (c) and (d) without consideration of the short-wavelength phonons that are cut off by the finite element shape functions.

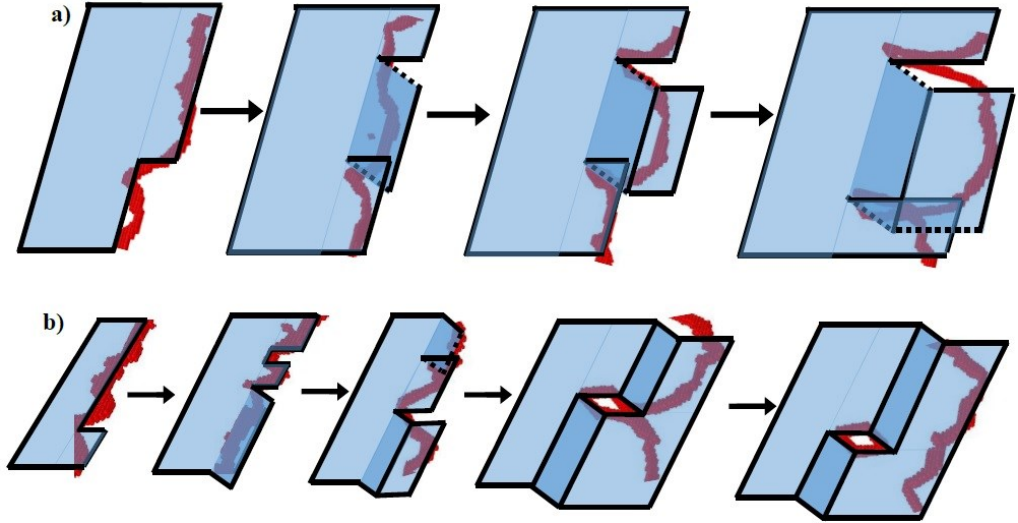

Fig. 8 Time sequences of a segment of the dislocation line to illustrate the atomistic process of (a) self-pinning and (b) debris loop formation that result from cross-kinks of a 2968-nm-long dislocation line.

## C. Heteroepitaxial growth of PbTe on PbSe (111)

To test the accuracy of CAC for finite-temperature simulation of epitaxial growth, we compare the CAC simulation results of the growth processes of PbSe on PbTe(111) and PbTe on PbSe(001) with MD simulations of the same models. A small substrate with dimensions of 50nm×50nm×10nm is used for both systems in order for MD to simulate the kinetic processes. In Fig. 9(a) and Fig. 10(a), we compare the CAC and MD simulation results of the dislocation densities per interface area (cm$^{-1}$) at different stages of growth for the PbSe/PbTe(111) and PbTe/PbSe(001) systems, respectively. In Fig. 9(b) and 10(b), we present the misfit dislocation networks at different epilayer coverages at the PbSe/PbTe(111) and PbTe/PbSe(001) interfaces, respectively. The dislocations are visualized using OVITO DXA (Dislocation Extraction Algorithm)[51].

As can be seen from Fig. 9 and Fig. (10), the CAC simulation results of the dislocation density and dislocation network agree well with those obtained by MD. This agreement demonstrates that, with significantly reduced DOFs, CAC can reproduce the kinetic processes of the formation of dislocations in PbSe/PbTe(111) and PbTe/PbSe(001) heteroepitaxial systems obtained by MD. Please see [52] for more simulation results of the PbTe/PbSe epitaxy.

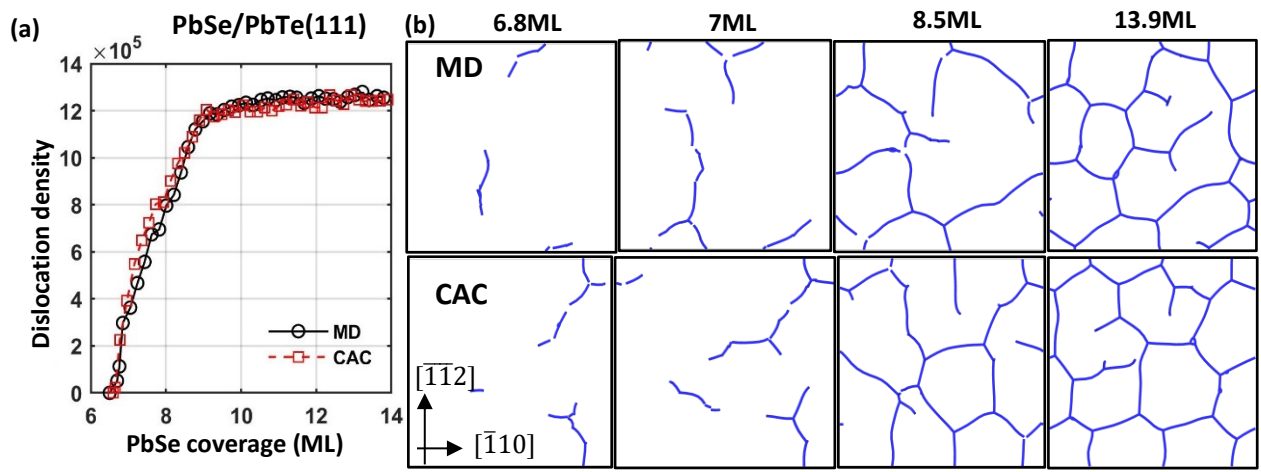

Fig. 9. A comparison between CAC and MD simulation results of (a) dislocation density and (b) dislocation networks in an PbSe/PbTe(111) epitaxial structure as a function of epilayer coverages.

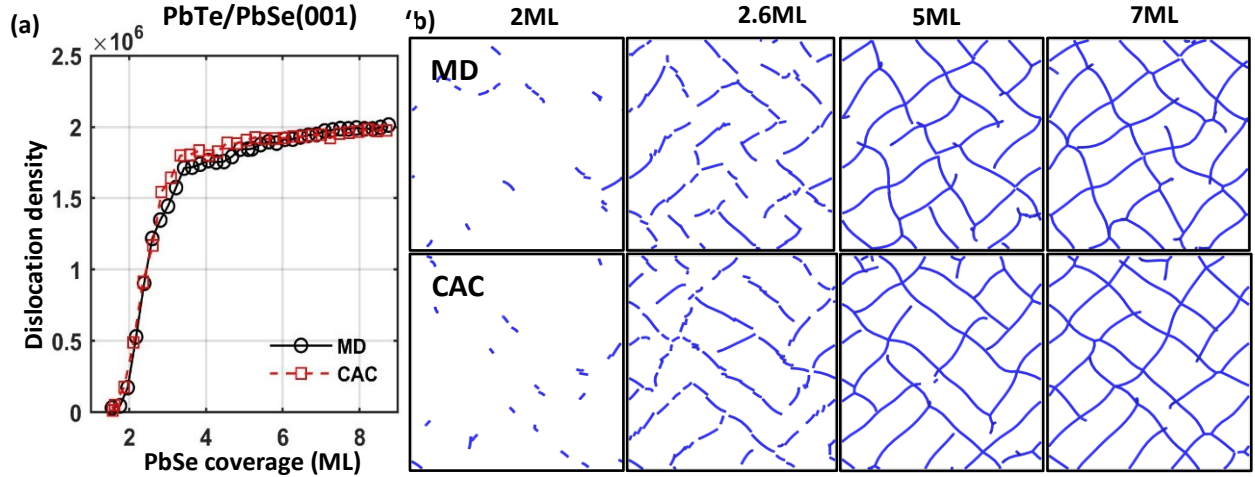

Fig. 10. A comparison between CAC and MD simulation results of (a) dislocation density and (b) dislocation networks in an PbTe/PbSe(001) epitaxial structures as a function of epilayer coverages.

## D. Attenuation of longitudinal acoustic phonons in silicon

Attenuation of hypersonic waves was experimentally observed more than six decades ago in perfect quartz crystals[53,54], and was theoretically explained to be caused by the interaction of the long wavelength phonons, i.e., sound waves, with short-wavelength thermal phonons[55-57]. The phenomenon was reproduced for higher frequency acoustic waves using picosecond ultrasonic measurements. Attenuation of 50 -100 GHz (100 GHz corresponds to an acoustic phonon with wavelength 90 nm) longitudinal acoustic phonons was observed in a thin Si wafer in the temperature range 50- 300 K[58]. Recently, using femtosecond laser pulses to generate and detect acoustic phonons, the propagation and attenuation of 1–1.4 THz (1.2 THz corresponds to phonon wavelength 6.7 nm) coherent longitudinal acoustic phonons were observed in GaN at room temperature[59].

Attenuation of long-wavelength acoustic waves is a manifestation of anharmonic interactions of the long wavelength phonons with short-wavelength thermal phonons[59]. To test the formulation in reproducing the interaction between a long-wavelength acoustic phonon with short-wavelength thermal phonons, the propagation of a phonon wave packet that centers at wavelength of 40nm in Si is simulated at room temperature. A comparison between CAC and MD simulation results of the evolution of the kinetic energy of the wave packet in space and in time is presented in Fig.11 (a) and Fig. 11(b), respectively. It is evident that the effect of the short wavelength phonons (thermal phonons) is captured in the CAC simulation and is in good agreement with that captured by atomically resolved MD simulation.

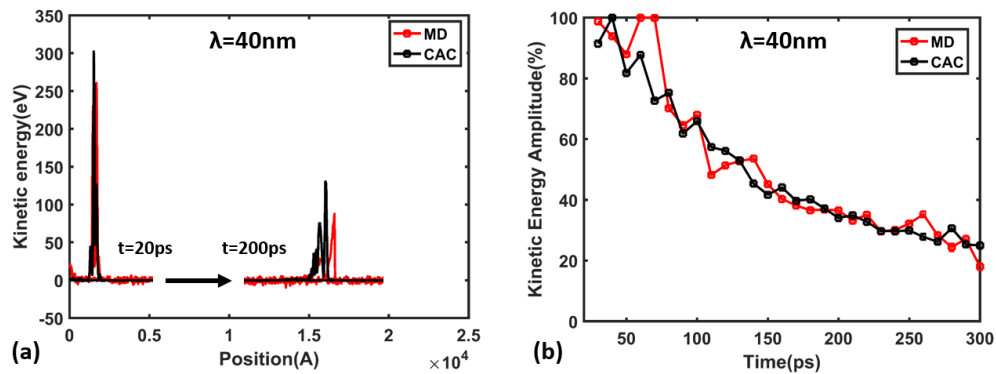

Fig. 11 Comparison between CAC and MD simulation results of the evolution of kinetic energy of an ultrashort phonon pulse in space and time, showing the attention of the waves due to the scattering by short wavelength phonons.

## VII. SUMMARY AND DISCUSSIONS

Classical field theories of mechanics, as advocated by Truesdell, Eringen, etc., were formulated top down based on a global form of conservation laws and an axiomatic constitutive theory, with the molecular or crystal structures of the materials being ignored. This work introduces a formalism that bottom up derives conservation equations from the classical atomistic model using the mathematical theory of distributions. Consequently, it provides an analytical link between atomistic and continuum descriptions of conservation laws and fluxes. It also significantly expands the scope, approach, and applicability of continuum mechanics, naturally leading to a concurrent multiscale methodology and simulation tools[60-65]. Fundamental differences between the bottom-up formulation and the top-down formulated field theories of mechanics may be summarized as follows.

(1) In distinction from that in classical field theories, the conservation equations and field quantities in this work are formulated based on a two-level structural description of crystalline materials, with both the atomic force field and the crystal structure, i.e., atomic arrangements within a unit cell, being built in the formulation. Consequently, the formulation is applicable to multiple length and time scales. It also eliminates the need of axioms, empirical rules, and assumptions for the constitutive theory in classical field theories. Moreover, it enables a field theory to exactly represent the underlying atomistic model.

(2) Different from that in classical continuum mechanics, thermodynamics, and many other multiscale theories or methods that define temperature as a basic quantity, temperature in this formulation is defined as a derived quantity in terms of the kinetic energy, as that in molecular kinetic theory and atomistic simulations, leading to a unified atomistic and field description of temperature and conservation laws at finite temperatures. The new finite-temperature linear momentum equation has the form of generalized Langevin equation and represents the underlying atomistic system in a statistical sense, as a result of the statistical nature of temperature field. Most importantly, the conservation equation of linear momentum completely governs thermal and/or mechanical processes from the atomistic to the continuum. It can be used to solve atomistic trajectories of systems at finite temperatures, as well as the evolution of field quantities in space and time, with atomic or multiscale resolution.

This work has also shown that the temperature field can be uniquely determined in a coarse-grained model only when the velocity field can be uniquely defined, such as for systems at thermal equilibrium or steady state. For such systems, temperature can be defined consistently across scales from the atomistic to the continuum, thereby extending the Concurrent Atomistic-Continuum (CAC) method to multiscale simulations of finite temperature processes.

It should be noted that materials processes at finite temperatures are complex, and that the formulation presented in this work only represents a first step towards addressing the complex problem of finite-temperature systems or processes. To fully address the length and time scale challenge in predictive simulation of materials and structures, we must bottom-up build a conceptual and mathematical framework for nonequilibrium mechanics of continua, which will be the goal of our future research.

**Acknowledgement**

This work is based on research supported by the US National Science Foundation under Award Number CMMI-2054607. We are grateful to Dr. Siddiq Qidwai for supporting the theoretical effort. The computer simulations are funded by the Advanced Cyberinfrastructure Coordination Ecosystem: Services & Support (ACCESS) allocation TG-DMR190008. We would like to thank Dr. Zexi Zheng, Dr. Rigelesaiyin Ji, and Dr. Yang Li for producing the computational/simulation results presented in this work.